\documentclass[
preprintnumbers,
nofootinbib,
 amsmath,amssymb,
 prd,aps,superscriptaddress,twocolumn,
floatfix,
]{revtex4-2}

\usepackage{graphicx}
\usepackage{amsmath}
\usepackage[caption=false]{subfig}
\usepackage{siunitx}
\usepackage{placeins}
\usepackage{color}
\usepackage{standalone}
\usepackage{dcolumn}
\usepackage{tensor}
\usepackage{bm}
\usepackage{microtype}
\usepackage{etoolbox}
\usepackage{amssymb}
\usepackage{mathrsfs}
\usepackage{accents}
\usepackage[normalem]{ulem}
\usepackage{soul} %
\usepackage[dvipsnames]{xcolor}
\usepackage[colorlinks,urlcolor=NavyBlue,citecolor=NavyBlue,linkcolor=NavyBlue,pdfusetitle]{hyperref}
\usepackage[all]{hypcap}
\usepackage[inline]{enumitem}
\usepackage[utf8]{inputenc}
\usepackage{xspace}
\usepackage[printonlyused, nolist]{acronym}
\usepackage{lineno}

\newcolumntype{a}{>{\columncolor{Gray}}c}
\newcolumntype{b}{>{\columncolor{white}}c}

\renewcommand{\vec}[1]{\mathbf{#1}}

\makeatletter
\newcommand*{\balancecolsandclearpage}{\close@column@grid \cleardoublepage \twocolumngrid}
\makeatother

\acrodef{LSC}[LSC]{LIGO Scientific Collaboration}
\acrodef{BH}{black hole}
\acrodef{NS}{neutron star}
\acrodef{PN}{Post-Newtonian}
\acrodef{BBH}{binary black-hole}
\acrodef{MBHB}{massive black-hole binary}
\acrodefplural{MBHB}[MBHBs]{massive black-hole binaries}
\acrodef{BNS}{binary neutron-star}
\acrodef{NSBH}{neutron-star black-hole}
\acrodef{EOB}{effective-one-body}
\acrodef{NR}{numerical relativity}
\acrodef{SNR}{signal-to-noise ratio}
\acrodef{GW}{gravitational-wave}
\acrodef{PSD}{power spectral density}
\acrodef{aLIGO}{Advanced Laser interferometer Gravitational-Wave Observatory}
\acrodef{AZDHP}{aLIGO zero detuned high power density}
\acrodef{GR}{general relativity}
\acrodef{PE}{parameter estimation}
\acrodef{LAL}{LIGO algorithm library}
\acrodef{TPI}{tensor-product interpolant}
\acrodef{SVD}{singular value decomposition}
\acrodef{ODE}{ordinary differential equation}
\acrodef{PDE}{partial differential equation}
\acrodef{ROM}{reduced order model}
\acrodef{QNM}{quasi-normal mode}
\acrodef{LISA}{Laser Interferometer Space Antenna}

\def\araa{{Ann.~Rev.~Astron.~Astrophys.}}
\def\apj{{Astrophys.~J.}}

\def\mnras{{Mon.~Not.~R.~Astron.~Soc.}}

\newcommand{\JHU}{\affiliation{William H. Miller III Department of Physics and Astronomy, Johns Hopkins University, Baltimore, Maryland 21218, USA}}

\begin{document}

\title{Classifying the generation and formation channels of individual LIGO-Virgo-KAGRA observations from dynamically formed binaries}

\author{Andrea Antonelli}
\email{aantone3@jh.edu; andrico@hotmail.it}
\author{Konstantinos Kritos}
\author{Ken K.~Y.~Ng}
\author{Roberto Cotesta}
\author{Emanuele Berti}

\JHU


\date{\today}

\begin{abstract}

We address two important questions in gravitational-wave astronomy. What is the astrophysical formation scenario leading to black-hole binary mergers? Did some of the merging black holes form hierarchically through previous generations of mergers?
Leveraging fast-to-generate astrophysical simulations from the \texttt{rapster} code and a random forest algorithm, we develop a pipeline to accurately classify the most likely generation and formation scenario of dynamically formed BHs on an event-by-event basis.
We test our framework on four merger events with features suggesting a dynamical origin: the large total mass event GW190521, GW190412 (with large mass asymmetry), and two events with effective spins antialigned with the orbital angular momentum (GW191109 and GW200225).
Within the models we consider, and assuming these events to be formed dynamically, we find that one of the component black holes in GW190521 formed from a previous merger with high probability ($\gtrsim 85\%$).
GW190521, GW191109 and GW200225 are compatible with formation through three-body interactions, while the most likely formation channel for GW190412 are two-body captures.
We also rule out that GW191109 contains only first-generation black holes with a probability of $ 97\%$.
Our pipeline could be useful to identify the evolutionary path of individual GW observations once it is trained on more comprehensive sets of binary formation simulations.

\end{abstract}

\maketitle

\section{Introduction}

The growing catalog of transient gravitational-wave (GW) events from the LIGO-Virgo-KAGRA (LVK) collaboration has opened up the possibility to discern the evolutionary paths that lead to the formation of compact object binaries. The two most common scenarios for black hole (BH) binaries involve either field formation, or dynamical formation in dense stellar environments (see e.g.~\cite{LIGOScientific:2010nhs,Mandel:2021smh} for reviews). 
Each scenario may involve subchannels, such as common envelope evolution~\cite{Dominik:2012kk,Broekgaarden:2021efa}, stable mass transfer~\cite{Olejak:2022zee}, and chemically homogeneous evolution~\cite{Riley:2020btf} for isolated binaries in the field; or formation in young massive star clusters~\cite{DiCarlo:2019pmf,Banerjee:2017mgr,Banerjee:2019ute,Banerjee:2021wqh,Ziosi:2014sra,Rizzuto:2021atw,Sedda:2021abh}, globular clusters~\cite{Kulkarni:1993fr,PortegiesZwart:1999nm,Rodriguez:2016kxx,Samsing:2017xmd,Rodriguez:2019huv,Askar:2016jwt,Arca-Sedda:2018qgq}, nuclear star clusters~\cite{Miller:2008yw,OLeary:2008myb,Antonini:2016gqe,Hoang:2017fvh,Fragione:2018yrb,Mapelli:2021syv,Wang:2020jsx,Sedda:2020jvg}, open clusters~\cite{Kumamoto:2018gdg,Rastello:2021gvw,Rastello:2018elx}, and active galactic nuclei~\cite{Samsing:2022fxi,Tagawa:2019osr} in the dynamical case.
Moreover, a binary BH (BBH) that has formed dynamically in a collisional system may have been assembled in various dynamical subchannels, such as two-body~\cite{1989ApJ...343..725Q}, three-body~\cite{1993ApJ...403..271G}, or binary-single exchange interactions~\cite{1996ApJ...467..359H}.

One of the main goals of GW astrophysics is to extract information about formation scenarios from LVK data. Population synthesis codes for binaries formed in the field~\cite{COMPASTeam:2021tbl,Belczynski:2005mr,Breivik:2019lmt} and in dynamical environments~\cite{Askar:2016jwt,Rodriguez:2021qhl,Mapelli:2021gyv,Antonini:2019ulv,Sedda:2021vjh,Kritos:2022ggc} typically involve a large number of hyperparameters. The common approach is to predict the properties of the population of observable mergers from astrophysical simulations with a specified set of hyperparameters, and identify features that can help to infer the hyperparameters by comparison with LVK catalogs \cite{Zevin:2022bfa,Kimball:2020opk,Kimball:2020qyd}.

Predicting datasets to match observations has the advantage of giving broad insights into the stages of the formation of LVK binaries and their host environments. The necessary tuning of the initial conditions is however not amenable to hyperparameter inference. The inverse problem (inferring the hyperparameters from data) allows one instead to gain direct insight from observations.

This kind of hyperparameter inference is only possible when the underlying astrophysical simulations are very fast to generate, either because the astrophysical simulations are fast to evaluate, or through the use of emulators \cite{Wong:2022flg}.
In this paper, we follow the first approach. We use the \texttt{rapster} code~\cite{Kritos:2022ggc}, which can compute the dynamical evolution of BHs in clusters in a matter of minutes, while retaining information from the hyperparameters that describe the cluster evolution at various stages of the simulations. We train a classification algorithm on a host of such \texttt{rapster} datasets. Our goal is to use the individual masses and spins of the BBHs and compare with individual observation in the LVK catalog to quantify (in a statistical sense) whether the binary components formed hierarchically, and infer their most likely formation subchannel.

As \texttt{rapster} only describes dynamically formed BHs, at present we can only apply the model to observations of binaries that we assume are formed dynamically. For this reason we test the pipeline with real GW data from events that are likely to have a dynamical origin: GW190412 (with asymmetric masses)~\cite{LIGOScientific:2020stg,Gerosa:2020bjb}, GW190521 (with components in the mass gap)~\cite{LIGOScientific:2020iuh}, and two events with effective spins antialigned with the orbital angular momentum, GW191109 and GW200225~\cite{Fishbach:2022lzq}. However the approach we describe in this paper can be applied to other population synthesis codes (whether the binaries are formed in the field or dynamically), as long as the event catalogs are reasonably fast to generate. By including more information in the training stage, the present pipeline could be fruitfully applied to LVK data to discern hyperparameters of binary systems on an individual-event basis.

The paper is structured as follows. In Sec.~\ref{sec:method} we describe the \texttt{rapster} datasets. In Sec.~\ref{sec:taining_validation} we discuss the classifier's underlying model and validation. In Sec.~\ref{sec:application} we apply the model to selected events from the LVK catalog, and we discuss our main results. In Sec.~\ref{sec:conclusions} we present some conclusions and directions for future work.

\section{Simulations}
\label{sec:method}

In this section we describe the model and code used to perform our simulations, as well as the initial conditions for the population of clusters we simulate (Sec.~\ref{sec:the_model}). Then we establish some notation for hierarchical mergers (Sec.~\ref{sec:hierarchical_mergers}) and dynamical BBH formation channels (Sec.~\ref{sec:dynamical_formation_channels}) that will be useful in the rest of the paper.

\subsection{The model}
\label{sec:the_model}

We use {\tt rapster} to generate the simulations needed to train our pipeline. Rapster is a population synthesis code for dynamical formation of BBHs in star clusters. The code simulates the BH subsystem in the core of the cluster in a semi-analytic way. It uses the {\tt SEVN} code to compute BH masses given the initial mass function of stars assuming a metallicity value~\cite{Spera:2017fyx}, and the {\tt precession} code to estimate the properties (mass and spin) of the merger remnants~\cite{Gerosa:2016sys}. The {\tt rapster} code requires as input the initial conditions of a star cluster, such as the initial cluster mass, initial half-mass radius, metallicity, and the redshift of cluster formation. It also assumes a model for the spin of first-generation (or 1g) BHs, i.e., BHs that form from the collapse of massive stars. For this work, the relevant output from {\tt rapster} is the ``mergers'' file, which contains information about the dynamical mergers that occurred throughout the history of a single cluster simulation (see the documentation in Ref. \cite{Kritos:2022ggc}).

The initial conditions for our simulated populations of clusters are chosen as follows. Motivated by astronomical observations~\cite{2019ARA&A..57..227K}, we draw the initial cluster mass $M_{\rm cl,0}$ from a Schechter-type distribution of the form $p(M_{\rm cl,0})\propto M_{\rm cl,0}^{-2}\exp(-M_{\rm cl,0}/M_{\rm t})$ in the range $[10^4,10^7]M_\odot$. Given current uncertainties, we set the truncation mass $M_{\rm t}$ to $10^9,10^{6},10^{5.3}M_\odot$, a choice also adopted in Ref.~\cite{Rodriguez:2018rmd}. We also set the initial half-mass radius to 0.5, 1.0 or 2.0~pc. The initial central stellar density is set to the Plummer value: $\rho_{\star,0}=3M_{\rm cl,0}/(4\pi(r_{\rm h,0}/1.3)^3)$, while the initial galactocentric radius of the cluster is sampled between 0 and 20~kpc assuming a S\'ersic profile with S\'ersic index $n=1$ and scale radius of 1~kpc. We sample the redshift of cluster formation from the Madau-Fragos distribution~\cite{Madau:2016jbv}, and for the metallicity we use the fit in Eq.~(6) of the same reference. Finally, we sample the spin of 1g BHs from a uniform distribution U$[0,\chi_{\rm max}]$, with $\chi_{\rm max}\in\{0.0,0.2,0.5\}$. All other initial conditions for {\tt rapster} are set to their default values. These choices result in a grid of $3\times3\times3=27$ simulations for the cluster population. However, for illustrative purposes we chose to focus on only 7 of these models, that will be discussed in the figures below and in Table~\ref{table:simulations0}. We denote these models by $S_x M_y r_z$, where the three subscripts $(x\,, y,\, z)$ refer to the values of $\chi_{\rm max}$, $\log_{10}M_{\rm t}$, and $r_{\rm h,0}/{\rm pc}$, respectively. The reason for this choice is to explore the sensitivity of our results to the choice of initial conditions of our cluster population, and not to exhaustively simulate every possible astrophysical situation.

\subsection{Hierarchical mergers}
\label{sec:hierarchical_mergers}

After the merger of two BHs, the remnant typically has a nonvanishing recoil velocity (or ``kick'') due to asymmetric GW emission~\cite{1983MNRAS.203.1049F}. If the GW kick does not exceed the escape velocity of the host cluster, then the merger remnant is retained in the cluster, and it can potentially merge again. We will refer to the merger remnant of two 1g BHs as a ``second generation'' (or 2g) BH. The 2g BH, if retained, may merge with either a 1g or another 2g BH. Thus we broadly identify three categories of mergers: ``1g+1g'', ``1g+2g'', and ``2g+2g''. In principle, it is possible that mergers involving even higher generation BHs could occur. However we find that the fraction of events with the next most popular flavor (the ``1g+3g'' mergers) only represent $\lesssim0.1\%$ of our populations. Thus, in this study we will neglect 3g or higher-generation BHs.

\subsection{Dynamical formation subchannels}
\label{sec:dynamical_formation_channels}

BBHs may form in {\tt rapster} in one of the following three ways:

\begin{itemize}
  \item[(i)] Three-body binary formation (3bb): the close interaction of three single BHs can result in the formation of a BBH if the third BH extracts enough kinetic energy from the system.
  \item[(ii)] Dynamical capture (cap): if the GW energy released during the hyperbolic encounter of two single BHs exceeds the kinetic energy of the two-body system at infinity, this may result in a highly eccentric and compact BBH.
  \item[(iii)] Exchanges (exc): the formation of a BBH occurs through a pair of exchange interactions of the form star-star$\to$BH-star$\to$BH-BH pair, where each BH substitutes each stellar component in succession. 
\end{itemize}

We denotes merger occurring from these three formation channels by ``3bb'', ``cap'', or ``exc'', respectively. We also include von Zeipel-Lidov-Kozai (ZLK) mergers, which involve the formation of hierarchical triples from BBH-BBH interactions. In the ZLK channel, the inner binary merges as a result of angular momentum and inclination oscillations in the orbit. This channel can lead to values of the eccentricity close to unity~\cite{Miller:2002pg}.

\section{Model training and validation}
\label{sec:taining_validation}

We train a machine-learning algorithm with the datasets from \texttt{rapster} described above. The goal is to (i) approximate a function $\vec y_{(i)} = f(\vec X_{(i)})$ with predictors $\vec{X}_{(i)}$ that can be readily linked to samples from GW observations, and (ii) predict variables $\vec y_{(i)} = \{y_{\text{gen}}, y_{\text{fc}}\}$ corresponding to the binary components' generation and to the formation subchannels. The full \texttt{rapster} datasets contain {19 cluster parameters and 9 BBH} parameters, but not all of them are necessary to accurately train the function. By trial and error and inspection of the correlation matrix in search for high correlations, we find that the component masses $m_{1,2}$ and dimensionless spin parameters $\chi_{1,2}$ are the most important parameters to include in the training datasets.
Both sets of parameters are in fact measured by the LVK collaboration, which makes the connection to observations direct.

The predictor set $\vec X_{\text{gen}}=\{m_1,m_2,\chi_1,\chi_2\}$ is sufficient to train accurate models for the binary components' generations. For the formation channel, we {reparameterize} the set to $\{\log M_{\rm tot},q,\chi_1,\chi_2\}$ to aid the numerical fitting. Here $M_{\rm tot}=m_1+m_2$ is the total mass of the binary, and $q=m_1/m_2\geq 1$ is the mass ratio. We find that two more hyperparameters  must be added to the predictor set for $y_{\text{fc}}$: the cluster formation redshift $z_{\rm cl}$ and initial cluster mass $M_{\rm cl}$. The cluster mass $M_{\rm cl}$ anticorrelates most strongly with $y_{\text{fc}}$, with Pearson coefficient $\sigma_{y_{\text{fc}}, M_{\rm cl}}=-0.35$, while $z_{\rm cl}$ correlates most strongly with the log of the total mass, with $\sigma_{\log M_{\rm tot}, M_{\rm cl}}=0.21$. The predictor set then reads $\vec X_{\text{fc}}=\{\log M_{\rm tot},q,\chi_1,\chi_2,z_{\rm cl}, M_{\rm cl}\}$.

At this stage we need an appropriate algorithm for the classification problem. We decided to choose Random Forest algorithms for their interpretability, as they can be thought as ensemble averages of decision trees. To avoid overfitting, we make sure that the individual decision trees do not ``grow too large'' by requiring each tree to have at most three nodes. We use the implementation from \texttt{sklearn}, and train our models on 70$\%$ of a randomly selected subset of \texttt{rapster} simulated binaries (with sample size $N\sim 10^5$). We compare the predictions of this emulator against a test set consisting of the remaining 30$\%$ of the binaries. For a reference training set $S_{0.0}M_{9.0}r_{1.0}$, we find that the pipeline can accurately identify all 28252 ``1g+1g'', 1444 ``1g+2g'', and 295 ``2g+2g'' test binaries, while correctly identifying the formation sub-channel in 74\% of the cases. Similar numbers are obtained when varying the training set.

Such a remarkable accuracy for the binary's generations should not be a surprise, as \texttt{rapster} clearly separates the various configurations of parent generations in the $(\chi_1,\,\chi_2)$ plane. The decision boundaries drawn by the classifier are such that for $\chi_1 \lesssim 0.5$ and $\chi_2 \lesssim 0.5$, the classifier predicts 1g+1g configurations; for $\chi_1 \gtrsim 0.5$ and $\chi_2 \gtrsim 0.5$, it predicts 2g+2g; and it predicts 1g+2g otherwise. The $\chi_1$ and $\chi_2$ LVK samples from individual events mostly populate one of these distinct regions, which helps in correctly predicting the test binaries' generation configuration. When the spin alone is not sufficient for a correct prediction, the fact that 1g+1g, 1g+2g and 2g+2g binaries populate distinct (though more overlapping) regions of the $(m_1,\,m_2)$ parameter space improves the model's accuracy. In particular, first-generation BHs cannot populate the high-mass region because the pair-instability supernova mechanism gives rise to an upper mass gap~\cite{Woosley:2021xba}. The exact value of the lower end of this upper mass gap depends on the details of the remnant-mass prescription model~\cite{Fryer:2011cx,Spera:2017fyx,Mandel:2020qwb}, on the $\rm ^{16}C(\alpha,\gamma) ^{16}O$ nuclear reaction rate~\cite{2021MNRAS.501.4514C,Farmer:2019jed,deBoer:2017ldl}, and on metallicity. With our choice of metallicities, and adopting the {\tt SEVN} model to compute the single-BH mass spectrum, the gap corresponds to $m_1\gtrsim55M_\odot$. However, the mass gap can be polluted by repeated BH mergers inside the cluster. The 1g+2g and 2g+2g populations are well separated because of the pairing prescriptions in {\tt rapster}~\cite{Kritos:2022ggc}. The mass ratio of 1g mergers in the model peaks near unity. Since the mass of a 2g BH is typically twice as large as that of a merging 1g BH, the mass ratio of 1g+2g events peaks at $q\simeq 0.5$, and that of 2g+2g events peaks at $q\simeq 1$. All of these features contribute to the accuracy with which our models predict the generation of the binary components.

\begin{figure}[t!]
\captionsetup{justification=raggedright,singlelinecheck=false}
\centering
\centerline{\includegraphics[width=\linewidth]{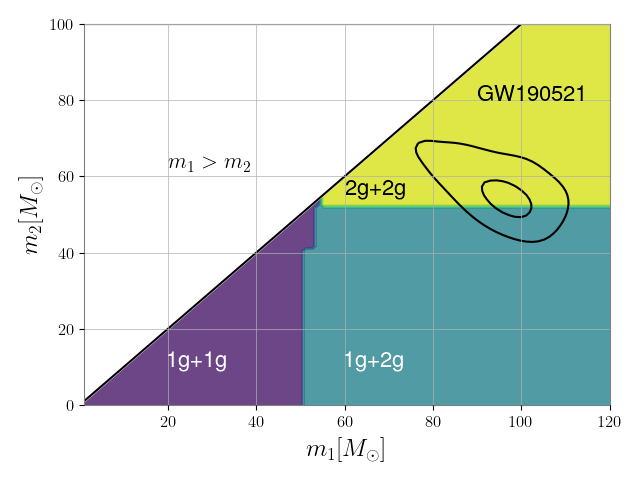}}
\caption{We plot the partitioning of the $(m_1,\,m_2)$ parameter space by the binary configurations under the fiducial model $S_{0.0}M_{9.0}r_{1.0}$. We also show the $1\sigma$ and $2\sigma$ contour plots for GW190521~\cite{LIGOScientific:2019lzm}. By using the event samples as input in the fit of Sec.~\ref{sec:method}, we can statistically establish the most likely binary configuration for GW190521.
}
\label{fig:GW190521_posterior_vs_sim}
\end{figure}

\section{Application to GW data}
\label{sec:application}

The models fitted in the previous section require fixed values as inputs, and thus most naturally apply to data analysis approached within a frequentist mindset. In this context, errors can be calculated by repetitions of the experiment. Most inference studies by the LVK collaboration are instead Bayesian, and the credible region associated to the measured parameters is obtained through a Markov Chain Monte Carlo (MCMC) analysis. In this work we reconcile the difference in mindsets by treating the samples defining the MCMC credible intervals as individual experiment repetitions, and inputting them as fixed values in the fits. 

Our approach  to connect cluster simulations with LVK data is exemplified in Fig.~\ref{fig:GW190521_posterior_vs_sim} for the case in which we want to predict the binary configuration. Here we show the partitioning by the classifier's decision boundaries, restricting the focus on the $(m_1,\,m_2)$ parameter space for illustrative purposes. We also plot the 1$\sigma$ and $2\sigma$ contours for the GW190521 event, chosen here as an example. By inputting the event's samples into the fit, we can establish what binary configuration it most likely pertains to by checking how many samples fall within each decision boundary. To predict the formation sub-channel we follow a similar approach. In this case, we must further specify some values for $z_{\rm cl}$ and $M_\text{cl}$, both of which appear in $\vec X_{\text{fc}}$. We choose $ M_\text{cl} = \{10^4, 3\times 10^4, 10^5,10^6,10^7\}M_\odot$ and $z_{\rm cl} = \{0.5,1,1.5,2,3\}$. These values cover the parameter space of our population of star clusters, the majority of which form at $z_{\rm cl}\sim 2$ (see Sec.~\ref{sec:the_model}). The final probabilities are obtained by averaging over pairs of ($z_{\rm cl},M_\text{cl}$).
For each event, we repeat the procedure varying the model hyperparameters that specify the training dataset.

The models presented above are trained on datasets that only include binary formation in clusters. The probabilities are in fact conditioned on the hypothesis that the input samples pertain to a dynamically formed BH. This current limitation of the procedure can be bypassed by training the fits to data from other formation channels. However, there are reasons to believe that certain GW events have indeed formed dynamically. In the analysis below we will illustrate the power of the approach by focusing on these specific events. We will start by discussing GW190521, arguably the most plausible event of dynamical origin.

\begin{figure}[t]
\centering
  \centerline{\includegraphics[width=.9\linewidth]{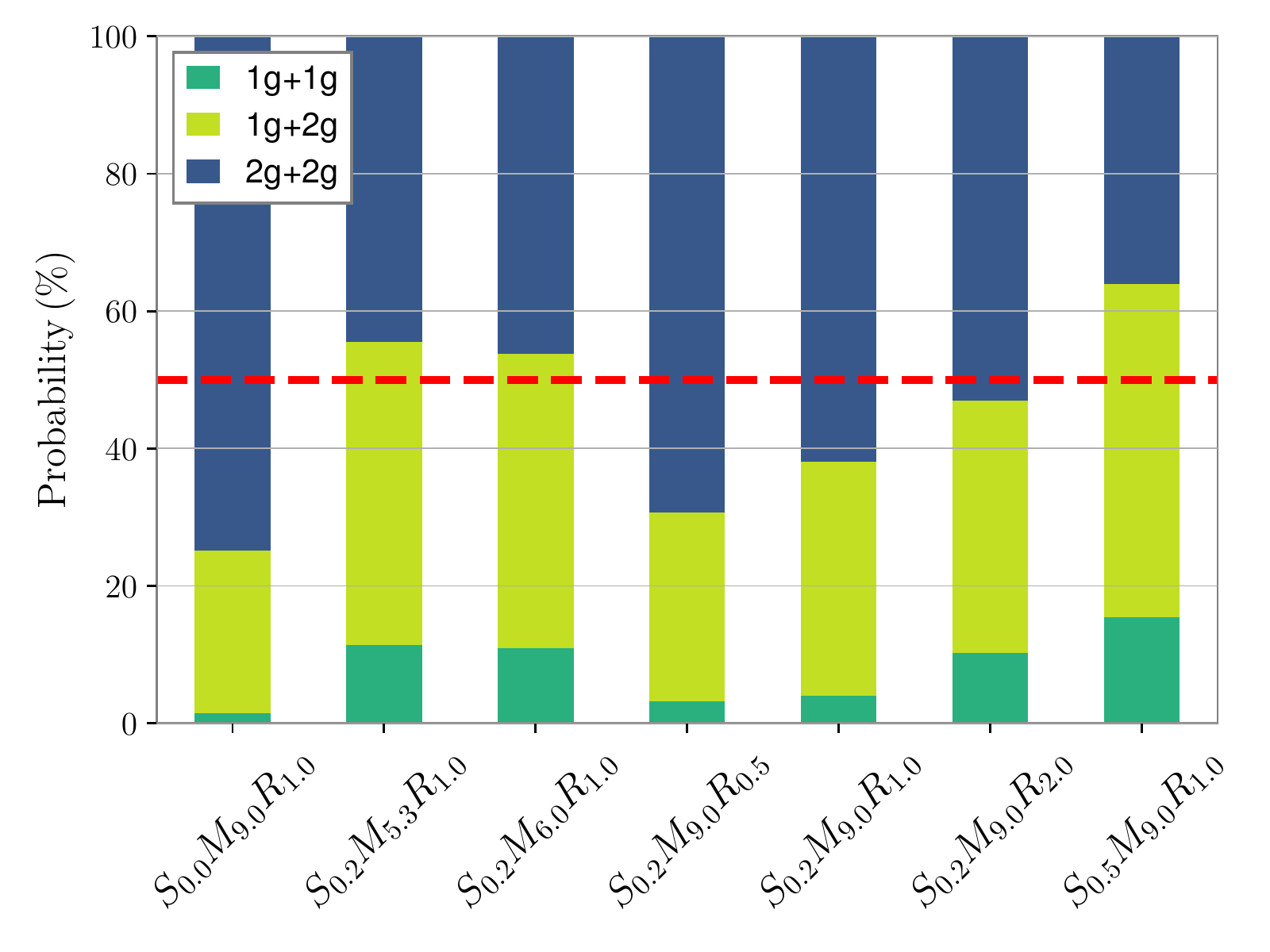}}
 \centerline{\includegraphics[width=.9\linewidth]{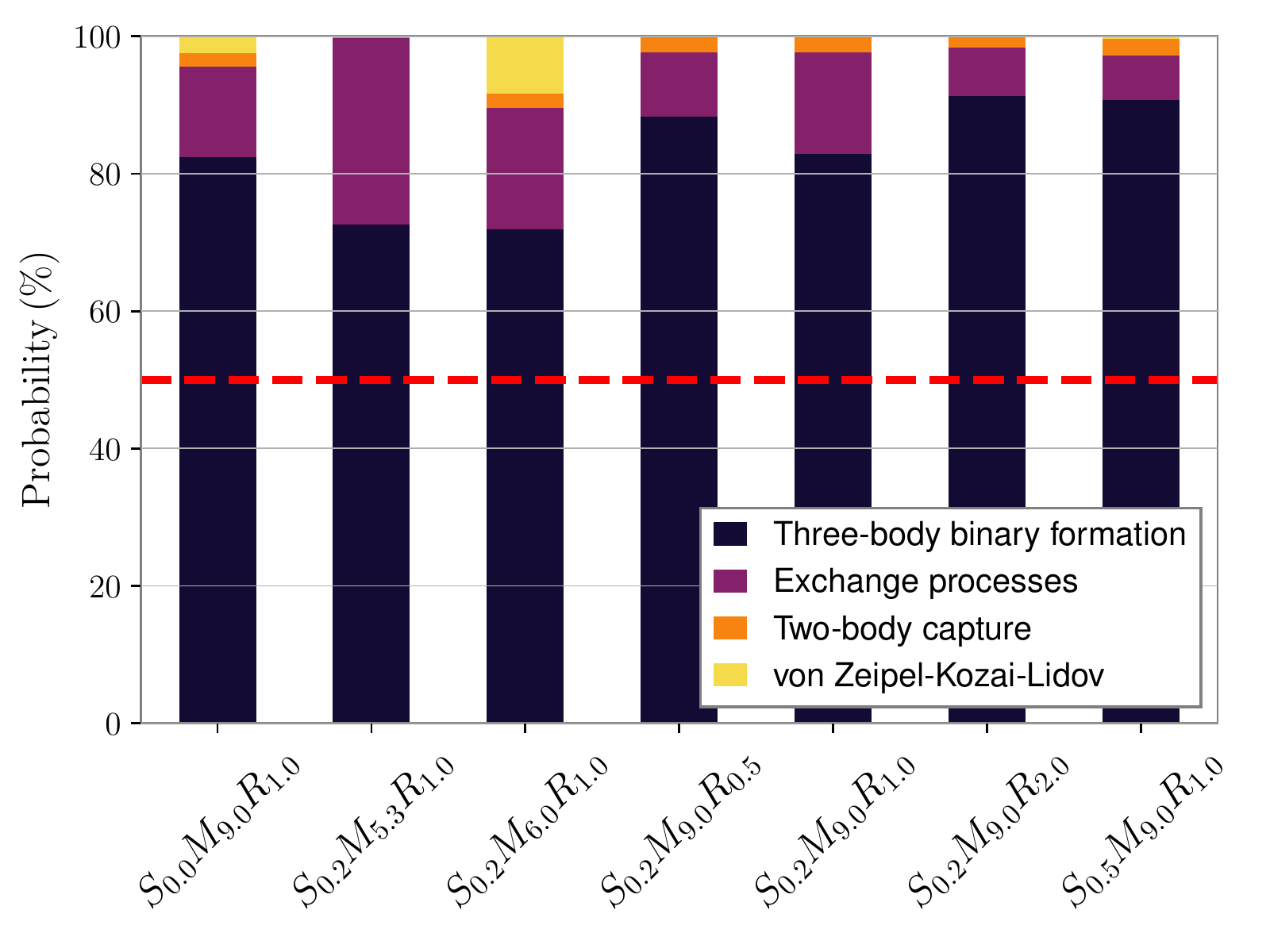}}
\caption{Top panel: probability that the GW190521 event is a 1g+1g, 1g+2g or 2g+2g binary. The binary is unlikely to contain only 1g BHs. Bottom panel: probability that it was formed through three-body interactions, exchange processes, two-body capture, or the ZLK mechanisms. Three-body interactions are the most likely subchannel among those considered in this paper. In each panel we show the 50\% probability as a dashed red line.}
\label{fig:test}
\end{figure}

\begin{table*}
	\caption{Classification results of the analysis for the GW190521, GW190412, GW191109 and GW200225 events. To guide the eye, we show in bold those binary configurations or formation subchannels that are most favored in our analysis, as well as the highest probabilities in each case we studied. We find GW190521 to be most likely composed of 2 first-generation BHs formed through 3 body interactions; GW190412 to be most likely a result of two-body captures (though with a smaller probability $\sim 50\%$); and  GW191109 and GW200225 to be most likely formed through three-body interactions. GW191109 is most likely a 2g+2g binary, with a very small chance that it contains a 1g BH. Our analysis is unable to correctly establish the generations of the BHs in GW190412 and GW200225, even though for the latter it seems plausible that no 1g BH is present in the binary.
		\label{table:simulations0}}
	\begin{ruledtabular}
		\begin{tabular}{c | ccc | cccc || ccc | cccc }
			
            & \multicolumn{6}{c}{\bf{ GW190521}}
			&\multicolumn{8}{c}{\bf{ GW190412}} 
			\\
   
            Model 
			&    $\hat p_\text{1g+1g}$ 
            &    $\hat p_\text{1g+2g}$ 
            &    {\bf $\hat p_\text{2g+2g}$}
            &    {\bf $\hat p_\text{3bb}$} 
            &    $\hat p_\text{exc}$ 
            &    $\hat p_\text{cap}$ 
            &    $\hat p_\text{ZLK}$
            &    $\hat p_\text{1g+1g}$ 
            &    $\hat p_\text{1g+2g}$ 
            &    $\hat p_\text{2g+2g}$  
            &    $\hat p_\text{3bb}$ 
            &    $\hat p_\text{exc}$ 
            &    {\bf $\hat p_\text{cap}$}
            &    $\hat p_\text{ZLK}$
            
			\\
	$S_{0.0}M_{9.0}r_{1.0}$ & 1$\%$ & 24$\%$ & \bf{75}$\%$ & \bf{82}$\%$ & 13$\%$ & 2$\%$ & 3$\%$ &   0$\%$ &  20$\%$ &  \bf{80}$\%$ &  43$\%$ &  21$\%$ &  \bf{54}$\%$ &  0$\%$ \\
	$S_{0.2}M_{5.3}r_{1.0}$ & 11$\%$ & 44$\%$ & \bf{45}$\%$ & \bf{73}$\%$ & 27$\%$ & 0$\%$ & 0$\%$ &  31$\%$ &  \bf{69}$\%$ &  0$\%$ &  40$\%$ &  1$\%$ &  \bf{58}$\%$ &  0$\%$ \\
	$S_{0.2}M_{6.0}r_{1.0}$ & 11$\%$ & 43$\%$ & \bf{46}$\%$ & \bf{72}$\%$ & 18$\%$ & 2$\%$ & 8$\%$ &  29$\%$ &  \bf{56}$\%$ &  0$\%$ &  39$\%$ &  0$\%$ &  \bf{61}$\%$ &  0$\%$ \\
	$S_{0.2}M_{9.0}r_{0.5}$ & 3$\%$ & 28$\%$ & \bf{69}$\%$ & \bf{88}$\%$ & 10$\%$ & 2$\%$ & 0$\%$ &  0$\%$ &  26$\%$ &  \bf{73}$\%$ &  34$\%$ &  2$\%$ &  \bf{64}$\%$ &  0$\%$ \\
	$S_{0.2}M_{9.0}r_{1.0}$ & 4$\%$ & 34$\%$ & \bf{62}$\%$ & \bf{83}$\%$ & 15$\%$ & 2$\%$ &  0$\%$  & 1$\%$ &  38$\%$ &  \bf{61}$\%$ &  35$\%$ &  0$\%$ &  \bf{65}$\%$ &  0$\%$ \\
	$S_{0.2}M_{9.0}r_{2.0}$ & 10$\%$ & 37$\%$ & \bf{53}$\%$ & \bf{91}$\%$ & 7$\%$ & 2$\%$ & 0$\%$  & 4$\%$ &  \bf{81}$\%$ &  15$\%$ &  38$\%$ &  0$\%$ &  \bf{61}$\%$ &  0$\%$ \\
	$S_{0.5}M_{9.0}r_{1.0}$ & 15$\%$ & \bf{49}$\%$ & 36$\%$  & \bf{91}$\%$ & 6$\%$ & 3$\%$ & 0$\%$ &  \bf{52}$\%$ &  46$\%$ &  2$\%$ &  36$\%$ &  0$\%$ &  \bf{64}$\%$ &  0$\%$ \\
			
		\end{tabular}
	\end{ruledtabular}
	\begin{ruledtabular}
		\begin{tabular}{c | ccc | cccc || ccc | cccc }
            & \multicolumn{6}{c}{\bf{ GW191109}}
			&\multicolumn{8}{c}{\bf{ GW200225}} 
			\\
            Model 
			&    $\hat p_\text{1g+1g}$
            &    $\hat p_\text{1g+2g}$ 
            &    {\bf $\hat p_\text{2g+2g}$}  
            &    {\bf $\hat p_\text{3bb}$}
            &    $\hat p_\text{exc}$ 
            &    $\hat p_\text{cap}$ 
            &    $\hat p_\text{ZLK}$
            &    $\hat p_\text{1g+1g}$ 
            &    $\hat p_\text{1g+2g}$ 
            &    $\hat p_\text{2g+2g}$  
            &    {\bf $\hat p_\text{3bb}$} 
            &    $\hat p_\text{exc}$ 
            &    $\hat p_\text{cap}$ 
            &    $\hat p_\text{ZLK}$
            
			\\
			$S_{0.0}M_{9.0}r_{1.0}$  &  0$\%$ &  14$\%$ &  \bf{86}$\%$ &           \bf{92}$\%$ &  7$\%$ &  0$\%$ &  1$\%$ &  2$\%$ &  25$\%$ &  \bf{73}$\%$ &           \bf{57}$\%$ &  3$\%$ &  40$\%$ &  0$\%$ \\
	$S_{0.2}M_{5.3}r_{1.0}$  &  2$\%$ &  37$\%$ &  \bf{61}$\%$ &           \bf{69}$\%$ &  31$\%$ &  0$\%$ &  0$\%$ &  12$\%$ &    \bf{88}$\%$ &  0$\%$ &           \bf{61}$\%$ &  9$\%$ &  30$\%$ &  0$\%$ \\
	$S_{0.2}M_{6.0}r_{1.0}$  &  2$\%$ &  36$\%$ &  \bf{62}$\%$ &           \bf{73}$\%$ &  21$\%$ &  0$\%$ &  5$\%$ &  11$\%$ &  \bf{49}$\%$ &  40$\%$ &           \bf{55}$\%$ &  4$\%$ &  41$\%$ &  0$\%$ \\
	$S_{0.2}M_{9.0}r_{0.5}$  &  0$\%$ &  19$\%$ &  \bf{81}$\%$ &           \bf{94}$\%$ &  5$\%$ &  1$\%$ &  0$\%$ &  4$\%$ &  30$\%$ &  \bf{66}$\%$ &           \bf{49}$\%$ &  13$\%$ &  38$\%$ &  0$\%$\\
	$S_{0.2}M_{9.0}r_{1.0}$  &  0$\%$ &  24$\%$ &  \bf{76}$\%$ &           \bf{85}$\%$ &  13$\%$ &  0$\%$ &  2$\%$ &  5$\%$ &  39$\%$ &  \bf{56}$\%$ &           \bf{53}$\%$ &  4$\%$ &  43$\%$ &  0$\%$\\
	$S_{0.2}M_{9.0}r_{2.0}$  &  2$\%$ &  30$\%$ &  \bf{68}$\%$ &           \bf{92}$\%$ &  8$\%$ &  0$\%$ &  0$\%$ &  6$\%$ &  \bf{55}$\%$ &  39$\%$ &           \bf{52}$\%$ &  3$\%$ &  43$\%$ &  2$\%$\\
	$S_{0.5}M_{9.0}r_{1.0}$  &  3$\%$ &  42$\%$ &  \bf{55}$\%$ &           \bf{91}$\%$ &  7$\%$ &  1$\%$ &  0$\%$ &  21$\%$ &  \bf{53}$\%$ &  26$\%$ &           48$\%$ &  3$\%$ &  \bf{49}$\%$ &  0$\%$\\
			
		\end{tabular}
	\end{ruledtabular}
\end{table*}

\subsection{GW190521}

The GW190521 BBH components have estimated masses  $m_1 = 85^{+21}_{-14} M_\odot$ and $m_2 = 66^{+17}_{-18} M_\odot$, hence both components have high probability of being within the upper mass gap~\cite{LIGOScientific:2020iuh}. This suggests that the binary could have formed dynamically through hierarchical mergers in clusters \cite{LIGOScientific:2020ufj, Zevin:2020gbd}.

In Fig.~\ref{fig:GW190521_posterior_vs_sim} we plot the LVK samples for this event over the simulation data points of model $S_{0.0}M_{9.0}r_{1.0}$. The binary generation classifier predicts probabilities of 1$\%$, 24$\%$ and 75$\%$ that the merger resulted from a 1g+1g, 1g+2g and 2g+2g binary, respectively. This quantifies what is visually apparent from Fig.~\ref{fig:GW190521_posterior_vs_sim}, i.e., that most configuration data points (and all those within $2\sigma$) are compatible with at least one 2g BH overlapping with the LVK samples. These conclusions can change as we vary the underlying model, as we show in Fig.~\ref{fig:test} (and in Table~\ref{table:simulations0}, where we list results for all of the events we consider).

In many cases the 2g+2g configuration has a probability larger than 50$\%$, but in some cases the probability drops to $\sim 30\%$ (top panel of Fig.~\ref{fig:test}). This suggests that we cannot make a robust prediction within the context of the $\texttt{rapster}$ models. For example, the 2g+2g probability is quite low for the model $S_{0.5}M_{9.0}R_{1.0}$, because in that case 1g BHs are born with higher spins. Higher spins result in large recoil velocities of the merger remnant, and therefore in lower retention probability of 2g BHs. Note, however, that there is still a high probability ($ \gtrsim 85\%$) that {\em at least one} of the merging BHs in GW190521 is 2g.

A similar analysis for the formation channels is shown in the bottom panel of Fig.~\ref{fig:test}. In this case, a comparison with the $\texttt{rapster}$ models implies that the most likely formation channel (with a probability $>70\%$) are three-body interactions. This is because GW190521 has a ``typical'' formation path in the \texttt{rapster} datasets, and 3bb is the most likely dynamical channel according to our \texttt{rapster} models.

\subsection{GW190412}

The GW190412 event exhibits a high asymmetry in the masses ($q\sim 1/3$) that points to a possible dynamical formation~\cite{LIGOScientific:2020stg,Gerosa:2020bjb}. However, running the event's samples through the pipeline described above leads to inconclusive results on the hierarchical character of the binary components. The predictions in Table \ref{table:simulations0} are very sensitive to changes in the underlying simulations: in some models the presence of a 1g BH is strongly disfavoured (with $\hat p_{\rm 1g+1g} \lesssim 1\%$), while in others it is the most likely scenario (with $\hat p_{\rm 1g+1g} \gtrsim 50\%$). This suggests that, under the assumptions used in the \texttt{rapster} models, slightly asymmetric mass ratios {\em alone} are not a good discriminant to determine whether one BH results from a previous merger. The loophole in this case are simulations of the ``$S_{0.5}M_yR_z$'' class. In these simulations, a smaller fraction of merger remnants is retained in the cluster, and the low mass ratio region is still dominated by 1g+1g mergers (see e.g. Fig.~11 in Ref. \cite{Kritos:2022ggc}). In these scenarios, the predictor will favor 1g+1g over 1g+2g despite the low mass ratio, simply because the 1g+1g rate is higher.

The formation subchannel analysis is more insightful. Across all the models we explored, the probability that the binary is formed through 2-body capture is $ \sim 60\%$.  Again, \texttt{rapster} can help us gain insight on the physics of the problem. 
In the capture scenario the lighter component is typically exchanged, but on a timescale much {\em longer} than the merger timescale. Exchanges typically lead to the ejection of the lightest component, and to a binary mass ratio closer to unity. In 2-body captures there is a small probability to equalize the masses, and the binaries tend to have some residual mass asymmetry. Conversely, in the second most favored scenario (3bb), exchanges that tend to equalize the masses happen on timescales {\em shorter} than the merger timescale, leading to less configurations exhibiting mass asymmetry. This explains why the 3bb scenario is disfavored, but not negligibly so. 

\subsection{GW191109 and GW200225}

Reference \cite{Fishbach:2022lzq} suggested that the GW191109 and GW200225 events may be formed through hierarchical mergers, as suggested by their mostly negative effective spins. Moreover, GW191109 is massive enough to suggest it was formed hierarchically. Our analysis excludes that GW191109 contains a 1g BH ($\hat p_{\rm 1g+1g}< 3\%$), with a slight preference for a 2g+2g configuration ($\hat p_{\rm 2g+2g}> 50\%$).

There is a strong preference for a 3bb formation channel for this event, with $\hat p_{\rm 3bb}$ ranging from $\sim 70\%$ to $90\%$. Again, this is consistent with our picture of cluster dynamics, where the heavier 2g BHs sink into the core of the cluster (if retained) and pair up with another BH through a three-body interaction.

Our predictions on the hierarchical origin of the GW200225 event are more inconclusive. In particular, the $S_{0.2}M_{5.3}r_{1.0}$ model appears to skew the predictions. Our cluster population in this case is dominated by low-mass clusters with smaller escape velocities, that do not easily retain their 2g products. The most likely formation channel is still the 3-body subchannel, although only marginally: we predict $\hat p_{\rm 3bb} \sim 50\%$, which is comparable to the probabilities for the two-body capture scenario.

This high sensitivity to the formation models is a reminder that detailed analyses with more comprehensive astrophysical models are needed to draw more definite conclusions on the evolutionary history of each of the GW events analyzed in this section.

\section{Conclusions}
\label{sec:conclusions}

Finding out the hierarchical origin and formation scenarios of BHs remains one of the key challenges in GW astrophysics. In this paper we have presented a pipeline that, leveraging machine-learning classification algorithms and fast-to-generate astrophysical simulations, can be applied to GW data to establish the most likely formation scenario of a given event, and the generation of the binary components. Our pipeline is trained on cluster simulations datasets from \texttt{rapster}. It uses mass and spin samples from GW observations as input, and returns key metrics for the binary evolutionary history as output.

We have applied the pipeline to GW190412, GW190521, GW191109, and GW200225, all of which have a reasonable probability (based on physical considerations) of being formed dynamically~\cite{LIGOScientific:2020ufj,Gerosa:2020bjb,Zevin:2020gbd,LIGOScientific:2020stg,Fishbach:2022lzq}.

Within the models we consider, and assuming that these events are indeed formed dynamically, we draw the following conclusions:

\begin{itemize}
    \item We exclude that GW190521 is composed of only 1g BHs $(p_{\rm 1g+1g}\lesssim 15\%)$, while it is likely to be formed through three-body interactions $(p_{\rm 3bb}\gtrsim 70\%)$.
    \item We find that GW190412 is formed through two-body captures with $p_{\rm cap} \sim 60\%$.
    \item We rule out that GW191109 contains only 1g BHs $(p_{\rm 1g+1g}\lesssim  3\%)$, with three-body interactions being the most likely scenario $(p_{\rm 3bb}\gtrsim 70\%)$.
    \item We find that GW200225 may result from three-body interactions with $p_{\rm 3bb}\gtrsim 50\%$.
\end{itemize}

Note that our method accounts for the relative rate of different classes of formation scenarios. For example, the classifier favored 1g+1g merger for GW190412 when we assumed a higher spinning population of 1g BHs, because in that case the tail of the 1g+1g distribution overcomes the relative rate of 1g+2g binaries in the low mass ratio regime. Moreover, the 3bb subchannel has a high probability of being associated to the events because it is typically the most likely mechanism for the formation of a BBH in a star cluster, even though there is a degeneracy where both subchannels populate roughly the same region in the $(m_1,\,m_2)$ parameter space.

As pointed out in Sec.~\ref{sec:hierarchical_mergers}, in this analysis we have neglected third- or higher-generation BHs, because they represent a small fraction of our population. However, since 3g BHs are typically heavier that 2g BHs, most mergers with a 3g component would populate the region above $m_1\sim70M_\odot$, that is currently underpopulated. Our fit is effectively an extrapolation when we consider LVK samples in those underpopulated regions. More simulation points and the inclusion of higher generation BHs would be required to probe the high-mass region with greater accuracy.

Our results depend on the astrophysical models used to train the classification algorithm, and (more specifically) on the choice of the \texttt{rapster} model to generate the cluster simulations. However, our approach should be considered as a proof of principle, and it is modular. Follow-up studies considering other simulation codes, such as \texttt{CMC} \cite{Rodriguez:2016kxx} and {\tt MOCCA} \cite{Askar:2016jwt}, are needed before drawing definite conclusions on the origin of the binaries we analyzed above. Our method is flexible enough to accommodate these improvements, as long as the underlying simulations are fast to generate. We also plan to extend the method to include field-formation scenarios. In this spirit, we have made the code publicly available online \cite{githuburl}. By coupling the classifier to a larger set of astrophysical simulations, it should be possible to draw more solid conclusions on the astrophysical origin of GW observations on an event-by-event basis.

\acknowledgments

We thank Floor Broekgaarden, Veome Kapil and Mike Zevin for very useful discussions.
The authors are supported by NSF Grants No. AST-2006538, PHY-2207502, PHY-090003 and PHY-20043, and NASA Grants No. 20-LPS20-0011 and 21-ATP21-0010.
KKYN is supported by a Miller Fellowship at Johns Hopkins University.
This research has made use of data or software obtained from the Gravitational Wave Open Science Center (gwosc.org), a service of LIGO Laboratory, the LIGO Scientific Collaboration, the Virgo Collaboration, and KAGRA. LIGO Laboratory and Advanced LIGO are funded by the United States National Science Foundation (NSF) as well as the Science and Technology Facilities Council (STFC) of the United Kingdom, the Max-Planck-Society (MPS), and the State of Niedersachsen/Germany for support of the construction of Advanced LIGO and construction and operation of the GEO600 detector. Additional support for Advanced LIGO was provided by the Australian Research Council. Virgo is funded, through the European Gravitational Observatory (EGO), by the French Centre National de Recherche Scientifique (CNRS), the Italian Istituto Nazionale di Fisica Nucleare (INFN) and the Dutch Nikhef, with contributions by institutions from Belgium, Germany, Greece, Hungary, Ireland, Japan, Monaco, Poland, Portugal, Spain. KAGRA is supported by Ministry of Education, Culture, Sports, Science and Technology (MEXT), Japan Society for the Promotion of Science (JSPS) in Japan; National Research Foundation (NRF) and Ministry of Science and ICT (MSIT) in Korea; Academia Sinica (AS) and National Science and Technology Council (NSTC) in Taiwan.

\bibliography{refs}

\end{document}